\documentclass[prd,aps,preprintnumbers,showpacs]{revtex4}
\usepackage{epsfig}
\def\VV{{\cal V}}
\def\AA{{\cal A}}

\newcommand{\be}{\begin{equation}}
\newcommand{\ee}{\end{equation}}
\newcommand{\bea}{\begin{eqnarray}}
\newcommand{\eea}{\end{eqnarray}}
\newcommand{\nn}{\nonumber}
\newcommand{\dd}{\displaystyle}
\def\ru{\rule[0.8truecm]{0.1truemm}{0.1truemm}}
\def\slash#1{\setbox0=\hbox{$#1$}#1\hskip-\wd0\dimen0=5pt\advance
\dimen0 by-\ht0\advance\dimen0 by\dp0\lower0.5\dimen0\hbox
to\wd0{\hss\sl/\/\hss}} 
\def\Black{}

\begin{document}
\begin{titlepage}


\preprint{BARI-TH 469/03} \preprint{DSF-2003/26 (Napoli)}

\title{
Charming penguins in $B\to K^*\pi,\,K(\rho,\omega,\phi)$ decays }

\author{\textbf{Claudia Isola}}
\affiliation{Dipartimento di Fisica, Universit\`{a} di Cagliari
and INFN,
Sezione di Cagliari, Italy\\
Centre de Physique Th{\'e}orique, {\'E}cole Polytechnique, 91128
Palaiseau Cedex, France}
\author{\textbf{Massimo Ladisa}}
\affiliation{Physics Department, Technion-Israel Institute of
Technology, Haifa 32000, Israel}
\author{\textbf{Giuseppe Nardulli}}
\affiliation{Dipartimento di Fisica dell'Universit{\`a} di Bari, Italy\\
Istituto Nazionale di Fisica Nucleare, Sezione di Bari, Italy}
\author{\textbf{Pietro Santorelli}}
\affiliation{Dipartimento di Scienze Fisiche, Universit{\`a} di
Napoli "Federico II", Italy\\
Istituto Nazionale di Fisica Nucleare, Sezione di Napoli, Italy}

\begin{abstract}
We evaluate the decays $B\to K^*\pi,\,K(\rho,\omega,\phi)$ adding
the long distance charming penguin contributions to the short
distance: Tree+Penguin amplitudes. We estimate the imaginary part
of the charming penguin by an effective field theory inspired by
the Heavy Quark Effective Theory and parameterize its real part.
The final results for branching ratios  depend on only two real
parameters and show a significant role of the charming penguins.
The overall agreement with the available experimental data is
satisfactory.
\end{abstract}


\maketitle
\end{titlepage}
\setcounter{page}{1}
\newpage
\section{Introduction} The CLEO II
\cite{Jessop:2000bv,Eckhart:2002mb},
 the BaBar \cite{Aubert:2001ap,Aubert:2001zf} and the
 Belle  \cite{Lu:2002qp} collaborations
 have recently reported data on the $B$ meson non leptonic decay  channels into a
 light vector and a pseudoscalar meson:
 \be B \rightarrow PV\ .\label{pv}\ee These data are of the foremost importance for the
 determination of the angles of the unitarity triangle. In particular, if one of the
 two mesons in the final state is a strange particle
the  decays (\ref{pv}) offer several channels for the extraction
of the $\gamma$ angle, thus providing alternatives  to the $K\pi$
decay mode. These non leptonic decays have been proposed long ago
\cite{Buras:1998rb,Gronau:1991dp,Fleischer:1998um,
Gronau:1994bn,Hernandez:1994rh,Gronau:1994rj,
Gronau:1995hm,Fleischer:1998bb,Gronau:1998an,
Neubert:1998pt,Neubert:1998jq,Neubert:1998re,Buras:2000gc} as a
method  for the extraction of the $\gamma$ angle. By this strategy
one relates the data to a theoretical amplitude given as a sum of
Tree and Penguin short distance contributions, whose relative weak
phase is indeed $\gamma$. The usual computational scheme is the
factorization hypothesis, either in the naive or in the QCD based
version \cite{Beneke:1999br},
\cite{Keum:2000ph,Keum:2000wi,Keum:2000ms,Lu:2000em,Kou:2001pm}.
Here one neglects non factorizable contributions as one proves
that they are of the order of $\Lambda/m_b$ and therefore
negligible in the $m_b\to\infty$ limit.

There is a class of contributions however that, although
suppressed in the infinite heavy quark mass limit, cannot be
neglected {\it a priori}. These are long distance contributions
known in the literature as charming penguin diagrams
\cite{Colangelo:1990gi,Ciuchini:1997hb,Ciuchini:2001gv,Isola:2001ar,
Isola:2001bn,Aleksan:2003qi}. Induced by the non leptonic
hamiltonian $\propto V_{cb}^*V_{cs}^{}{\bar
b}\gamma_\mu(1-\gamma_5) c\ {\bar c}\gamma^\mu(1-\gamma_5)s\,+{\rm
Fierz}\,+\,{\rm h.c.}$,
 they cannot contribute
in the vacuum saturation approximation. However, the insertion of
at least two charmed particles (charm+anticharm) between the
currents can produce sizeable contributions. As a matter of fact
the ${\cal O}(\Lambda\times m_b^{-1})$ suppression is compensated
by the Cabibbo-Kobayashi-Maskawa enhancement and the actual role
of these contributions can be established only as a result of some
dynamical calculations. In \cite{Isola:2001ar} and
\cite{Isola:2001bn} we estimated these contributions for the $B\to
PP$ decay channels and found that for the $K\pi$ case they are
indeed relevant. Their imaginary part can be evaluated with some
confidence using the heavy meson chiral effective theory corrected
for the light meson hard momenta. On the other hand, the real part
is less predictable; for example in \cite{Isola:2001ar} and
\cite{Isola:2001bn} the results strongly depend on the cutoff in
the loop integrals.

In the present paper we give an estimate  of the charming penguins
in the charmless  decays (\ref{pv}) for the cases $K^*\pi$ and
$K(\rho,\omega,\phi)$. We do not consider the particles
$\eta,\eta^\prime$ in the final state because the pseudoscalar
$SU(3)$ singlet is most probably contaminated by the glue and the
evaluation of these contributions cannot be reliably done within
our theoretical scheme. Due to the above mentioned difficulties we
compute by the effective theory only the imaginary part of the
amplitude, ${\cal I_{LD}}$. For the real part of the amplitude,
${\cal R_{LD}}$, we assume a simple parametrization ${\cal
R_{LD}}\propto{\cal I_{LD}}$, fixing the two  proportionality
constants (one for each of the SU(3)$_f$ multiplets comprising the
light vector mesons) by a fit. A similar calculation has been
recently attempted by \cite{Aleksan:2003qi}; these authors assume
a more phenomenological approach and parametrize the amplitude in
the more general way (two complex numbers), using $SU(3)$ flavor
symmetry to relate the different amplitudes. Therefore our
calculation can provide a dynamical test of the model in
\cite{Aleksan:2003qi}.

The outline of the paper is as follows. In Section \ref{sec.2}  we
estimate the tree diagram  contribution (or Short-Distance
contribution), computed in the factorization approximation. In
Section \ref{sec.3}  we estimate the absorptive part of the
charming penguin contribution. It is evaluated by use of the
effective lagrangian for light and heavy mesons based on the Heavy
Quark Effective Theory (for a review see
\cite{Casalbuoni:1996pg}). The main uncertainty of this approach
is the extrapolation to hard light meson momenta and in this
context we use an estimate of form factor given by us in
\cite{Isola:2001ar}. In Section \ref{sec.4} we present numerical
results for the branching ratios and the asymmetry. In our
previous papers we estimated the real part of the charming penguin
contributions by using a cut-off Cottingham formula; the results
where strongly cut-off dependent and for this reason we give here
a simple parametrization of the real part and we estimate the
relevant two real parameters by comparing with the data. The
overall result for both of the branching ratios and the
CP-asymmetries is rather satisfying and points to a significant
role of the charming penguins
 in the $B\to K^*\pi,\,K(\rho,\omega,\phi)$ decay modes.

\section{Short-Distance nonleptonic weak matrix elements\label{sec.2}}
The effective Hamiltonian for non-leptonic $B$ decays is given by
\begin{equation}
{\cal H}_{\rm eff} = {G_{F} \over \sqrt{2}} \left[V_{ub}^*
V_{us}(c_1 O_1^u + c_2 O_2^u) + V_{cb}^* V_{cs}(c_1 O_1^c + c_2
O_2^c) \nonumber \\
- V_{tb}^* V_{ts}\left( \sum_{i=3}^{10} c_i O_i + c_g O_g
\right)\right] \label{Heff}
\end{equation}
where $c_i$ are the Wilson coefficients evaluated at the
normalization scale $\mu = m_b$
\cite{Fleischer:1993gp,Fleischer:1994gr,Ciuchini:1993ks,Ciuchini:1994vr,Buras}
and next-to-leading order QCD radiative corrections are included.
$O_1$ and $O_2$ are the usual tree-level operators, $O_i$
($i=3,..., 10$) are the penguin operators and $O_g$ is the
chromomagnetic gluon operator. The ${\it c_i}$ in eq. (\ref{Heff})
are as follows \cite{Buras}:
$c_2=1.105,~c_1=-0.228,~c_3=0.013,~c_4=-0.029,~c_5=0.009,
~c_6=-0.033,~c_7/\alpha=0.005,~c_8/\alpha=0.060,
~c_9/\alpha=-1.283,~c_{10}/\alpha=0.266$. Moreover, for the
current quark masses we use the values
\begin{equation}
m_b=4.6\, GeV\;\;\;\;\;\; m_u\,=\,4\,MeV\;\;\;\;\;\;
m_d\,=\,8\,MeV\;\;\;\;\;\; m_s\,=\,0.150\,GeV.
\end{equation}
We define the $T$-matrix element
\be (2\pi)^4\delta^4(p_B-p_\pi-p_{K^*})\times{\cal
M}_{K^\ast\pi}=\langle K^\ast\pi |\, T\, | B \rangle \,, \ee
with a similar definition for the $K(\rho,\omega,\phi)$ final
state.  We separate short-distance and long-distance contributions
to the weak $B \to K^\ast \pi$ decay:
\begin{equation}
{\cal M} = {\cal M_{SD}} + {\cal M_{LD}} \label{ampl}\ ,
\end{equation}
and evaluate the short-distance part of the amplitude ${\cal
M_{SD}}$ using the operators in (\ref{Heff}) in factorization
approximation (with $ {\cal M_{SD}}(B \to f) \equiv- \langle f |\,
{\cal H}_{\rm eff}\, | B \rangle $ ). The results are in Tab.
\ref{tab:1}. Here ${\dd a_i = c_i + \frac{c_{i+1}}{3}}$ (i=odd)
and ${\dd a_i = c_i + \frac{c_{i-1}}{3}}$ (i=even). Moreover, if
$V$ is a vector meson, $P^{(\prime)}$ is a  pseudoscalar meson we
use the following definition for the matrix elements of weak
currents:
\be \label{leptonic} \langle P(p)|A_\mu|\, 0\, \rangle = -i\,
f_P\, p_\mu ~,~~~~~~~~~~~~\langle
V(\varepsilon,p)|V_\mu|\,0\,\rangle =
 f_V\, m_V\, \varepsilon^\ast_\mu\ ,
\ee
and
\bea \langle P^\prime(p^\prime)|V_\mu|P(p)\rangle & = & F_1(q^2)
\left[(p_\mu+p^\prime_\mu)-\frac{m_P^2-m_{P^\prime}^2}{q^2}\,
q_\mu\right]+ F_0(q^2)\frac{m_P^2-m_{P^\prime}^2}{q^2}\, q_\mu\\
<V(\epsilon,p')|V^\mu -A^\mu|P(p)>& =&\frac {2 V(q^2)} {M_P+M_V}
\epsilon^{\mu \nu \alpha \beta}\epsilon^*_{\nu} p_{\alpha}
p'_{\beta}\, -\, i (M_P+M_V) \epsilon^{*\mu}
 A_1 (q^2) \nn\\
& + & i \frac{(\epsilon^* \cdot q)}{M_P+M_V} (p+p')^\mu A_2
(q^2)\, +\, i (\epsilon^* \cdot q)\, \frac{2 M_V}{q^2} q^\mu\left[
A_3 (q^2)- A_0 (q^2)\right], \label{v1} \eea
where
\be A_3 (q^2)=\frac {M_V-M_P}{2 M_V} A_2(q^2) + \frac {M_V+M_P}
{2M_V} A_1(q^2) \label{v2} \ee
and $A_3(0)=A_0(0)$.
\begin{table}[ht!]
\caption{\label{tab:1} {\small Factorized ${\cal M_{SD}}$
amplitudes.}}
\begin{footnotesize}
\begin{center}
\begin{tabular}{|c|c|}
\hline
{\rm Process}  & ${\cal M_{SD}}$ \\
\hline
& \vspace{-0.18truecm} \\
$  {B}^{+}\, \to  K^{\ast\, 0} \pi^+$ & $ \dd + G_F\,\sqrt{2}\,
F_1^{B\to\pi}(m_{K^\ast}^2)\,f_{K^\ast}\, m_{K^\ast}\, V_{tb}^\ast
V_{ts}  \left [ a_4-\frac{ a_{10} }{ 2 } \right ]
\left(\varepsilon^*\cdot p_B\right) $ \\

& \vspace{-0.18truecm} \\
\hline
& \vspace{-0.18truecm} \\

$  B^{+}\, \to K^{\ast +}\pi^0$ & $\hspace{-1.0truecm}-G_F\,
m_{K^\ast}\, \Big \{F_1^{B\to\pi}(m_{K^\ast}^2)\,f_{K^\ast}\,
\left [ V_{ub}^\ast V_{us} a_2-V_{tb}^\ast V_{ts} (a_4+a_{10})\frac{}{}\!\!\right ]  $\\
& $ \hspace{2.0truecm}\dd +  A_0^{B \to K^\ast}(m_{\pi}^2) \,
f_\pi \, \left [V_{ub}^\ast V_{us} a_1+ V_{tb}^\ast V_{ts}\;
\frac{3}{2}
(a_7-a_{9})\right ]\Big\}\left(\varepsilon^*\cdot p_B\right)$ \\

& \vspace{-0.18truecm} \\
\hline
& \vspace{-0.18truecm} \\

$ B^0\, \to K^{\ast 0} \pi^0$& $\hspace{-1.5truecm} \dd
-\,G_F\,m_{K^\ast} \Big\{ A_0^{B\to K^\ast}(m_{\pi}^2)\, f_\pi
\left [ V_{ub}^\ast V_{us}\,  a_1 -
V_{tb}^\ast V_{ts}\, \frac{3}{2}(a_9-a_{7}) \right ] $ \\
& $\dd + F_1^{B\to\pi}(m_{K^\ast}^2)\,f_{K^\ast}\, V_{tb}^\ast
V_{ts} \; \left[a_4-\frac{a_{10}}{2} \right ] \Big
\}\left(\varepsilon^*\cdot p_B\right)$\\

& \vspace{-0.18truecm} \\
\hline
& \vspace{-0.18truecm} \\

$  B^0\, \to  K^{\ast +} \pi^-$ & $\dd -\, G_F\,\sqrt{2} \,
F_1^{B\to\pi}(m_{K^\ast}^2)\,f_{K^\ast}\, m_{K^\ast}\; \left [
\frac{}{} V_{ub}^\ast V_{us} \;  a_2\, -\, V_{tb}^\ast V_{ts}\;
\left
(\!\!\!\frac{}{}a_4+a_{10}\right ) \right ]\left( \varepsilon^*\cdot p_B\right ) $ \\

& \vspace{-0.18truecm} \\
\hline
& \vspace{-0.18truecm} \\

$  B^{+}\, \to K^{0} \rho^+$& $+\,G_F\,\sqrt{2}\,
A_0^{B\to\rho}(m_{K}^2)\,f_{K}\, m_{\rho} \; V_{tb}^\ast V_{ts}
\dd \left [ a_4-\frac{1}{2} a_{10}
-\frac{(2a_6-a_8)\, m_K^2}{(m_b+m_d)(m_d+m_s)} \right ]\left(\varepsilon^*\cdot p_B\right) $\\

& \vspace{-0.18truecm} \\
\hline
& \vspace{-0.18truecm} \\

$  B^{+}\, \to K^{+ }\rho^0$ & $\dd -\,G_F\, m_{\rho} \Big \{
A_0^{B\to\rho}(m_{K}^2)\,f_{K}\,
 \left [ V_{ub}^\ast V_{us}\, a_2
-V_{tb}^\ast V_{ts} \left (a_4+a_{10}-
\frac{2(a_6+a_8)\,m_K^2}{(m_b+m_u)(m_u+m_s)}\right )\right ]$\\
& $\hspace{-1truecm}\dd  + \, F_1^{B\to K}(m_{\rho}^2)\, f_\rho\,
 \left [V_{ub}^\ast V_{us}\, a_1 -  \frac{3}{2}\, V_{tb}^\ast V_{ts}\;
(a_7+a_{9})\right ]\Big \}\left(\varepsilon^*\cdot p_B\right)$\\

& \vspace{-0.18truecm} \\
\hline
& \vspace{-0.18truecm} \\

$ B^0\, \to K^{0} \rho^0$ & $\hspace{-2.4truecm}\dd -\, G_F\,
m_{\rho} \Big \{\, F_1^{B\to K}(m_{\rho}^2)\,f_\rho\, \left [
V_{ub}^\ast V_{us}\; a_1 - V_{tb}^\ast V_{ts}\;
\frac{3}{2}(a_7+a_9) \right ]$\\ \ru
 & $\hspace{2.0truecm}\dd +\, A_0^{B\to\rho}(m_{K}^2)\,f_{K}\,
V_{tb}^\ast V_{ts} \left[a_4-\frac{1}{2}a_{10}- \frac{(2
a_6-a_8)\, m_K^2}{(m_b+m_d)(m_d+m_s)} \right ] \Big \}
\left(\varepsilon^*\cdot p_B\right) $\\

& \vspace{-0.18truecm} \\
\hline
& \vspace{-0.18truecm} \\

$ B^0\, \to  K^+ \rho^-$ & $\dd -\, G_F\,\sqrt{2} \,
A_0^{B\to\rho}(m_{K}^2)\, f_{K}\, m_{\rho}\; \left [ \frac{}{}
V_{ub}^\ast V_{us} \;  a_2\, -\, V_{tb}^\ast V_{ts}\,
\left(a_4+a_{10}-\frac{2(a_6+a_8)\,
m_K^2}{(m_b+m_u)(m_u+m_s)}\right )
\right ]\left( \varepsilon^*\cdot p_B\right ) $ \\

& \vspace{-0.18truecm} \\
\hline
& \vspace{-0.18truecm} \\

$  B^+\, \to K^{+} \omega $ & $\dd\hspace{-3truecm} -G_F\,
m_{\omega}\, \left\{ F_1^{B\to K}(m_{\omega}^2)\, f_{\omega}
\left[ V_{ub}^\ast V_{us} a_1 -V_{tb}^\ast V_{ts}
\left(2(a_3+a_5)+\frac{1}{2}(a_7+a_9)\right) \right]\right.$\\
 & $\dd \hspace{1truecm}\left. + A_0^{B\to\omega}(m_K^2)\, f_K
\left[ V_{ub}^\ast V_{us}\, a_2 -V_{tb}^\ast V_{ts}
\left(a_4+a_{10}-\frac{2(a_6+a_8)m_K^2}{(m_b+m_u)(m_u+m_s)}\right)
\right]
\right\}\left(\varepsilon^*\cdot p_B\right) $\\

& \vspace{-0.18truecm} \\
\hline
& \vspace{-0.18truecm} \\

$  B^0\, \to K^0 \omega$ & $\dd\hspace{-2.5truecm} -G_F\,
m_{\omega}\, \left\{ F_1^{B\to K}(m_{\omega}^2)\, f_{\omega}
\left[ V_{ub}^\ast V_{us} a_1 -V_{tb}^\ast V_{ts}
\left(2(a_3+a_5)+\frac{1}{2}(a_7+a_9)\right) \right]\right .$\\
 & $\dd\left. -\, A_0^{B\to\omega}(m_K^2)\,f_K\,
V_{tb}^\ast
V_{ts}\left(a_4-\frac{1}{2}a_{10}-\frac{(2a_6-a_8)m_K^2}
{(m_b+m_d)(m_d+m_s)}\right)
\right\}\left(\varepsilon^*\cdot p_B\right) $\\

& \vspace{-0.18truecm} \\
\hline
& \vspace{-0.18truecm} \\

$  {B}^{+,0}\, \to  K^{+,0}\, \phi $ & $\dd + G_F\,\sqrt{2}\,
F_1^{B\to K}(m_{\phi}^2)\,f_{\phi}\, m_{\phi}\, V_{tb}^\ast V_{ts}
\left [ a_3+a_4+a_5-\frac{ a_7+a_9+ a_{10} }{ 2 } \right ]
\left(\varepsilon^*\cdot p_B\right)$\\
& \\
\hline
\end{tabular}
\end{center}
\end{footnotesize}
\end{table}

\section{Absorptive part\label{sec.3}}
\begin{center}
\begin{figure}[ht!]
\hskip 2cm \epsfxsize=12cm \epsfysize=7cm
\epsfbox{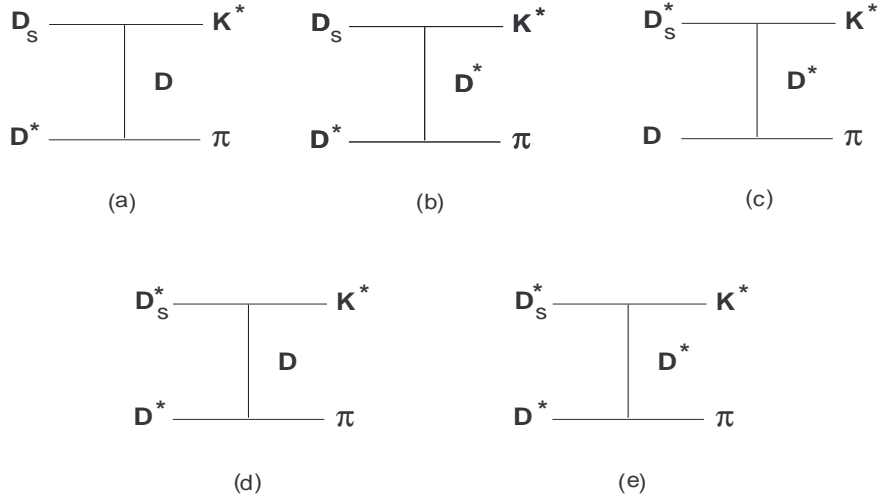} \vspace{3mm}
\caption{\label{fig:DDKpi}Diagrams for the calculation of the
amplitudes $D^{(*)}_s \,D^{(*)}\to K^*\pi$ amplitudes.}
\end{figure}
\end{center}

The computation of the discontinuity of the charming penguins
diagrams contributing to $B\to K^\ast\pi$ gives (cfr. analogous
diagrams in figure 2 of Ref. \protect\cite{Isola:2001ar})
\begin{eqnarray}{\rm Disc}{\cal M_{LD}} &=&
 2\, i\, {\rm Im}\,{\cal M_{LD}} =\, i\,(2\pi )^2 \int\frac{d^4
q}{(2\pi)^4}\delta_+(q^2-m^2_{D_s})\, \delta_+(p^2_{D^{(*)}}-m^2_D
)\times \cr &&\cr&\times &{\cal M}(B\to D_s^{(*)}D^{(*)})\, {\cal
M}(D_s^{(*)}D^{(*)}\to K^*\pi)\ =\cr&&\cr &=& \,i\,
\frac{m_D}{16\pi^2 m_B}\, {\sqrt{\omega^{*2}-1}}\, \int d{\bf n}
\, {\cal M}(B\to D_s^{(*)}D^{(*)})\, {\cal M}(D_s^{(*)}D^{(*)}\to
K^*\pi)\ ,
\end{eqnarray}
where the integration is over the solid angle and a sum over
polarizations is understood. A similar equation holds for the
$K(\rho,\omega,\phi)$ final state. The amplitudes for the decays
$B\to D_s^{(*)}D^{(*)}$ are computed by factorization and using
information on the Isgur-Wise function (see below). Diagrams for
the calculation of the amplitudes $D^{(*)}_s \,D^{(*)}\to K^*\pi$
are in Fig.\ref{fig:DDKpi}. A similar diagram can be drawn for the
$D^{(*)}_s \,D^{(*)}\to K(\rho,\omega,\phi)$ amplitudes.
The effective lagrangian to compute these diagrams can be written
as follows (see e.g. \cite{Casalbuoni:1996pg}):
\bea {\cal L} &=& i < H_b v^\mu D_{\mu ba} {\bar H}_a > + i g <H_b
\gamma_\mu \gamma_5 \AA^\mu_{ba} {\bar H}_a> \cr&+& i\beta
<H_bv^\mu\left(\VV_\mu-\rho_\mu\right)_{ba}{\bar H}_a> + i \lambda
<H_b \sigma^{\mu\nu} F_{\mu\nu}(\rho)_{ba} {\bar H}_a>~~.
\label{wise} \eea Here $<...>$ means the trace,\be
H=\frac{1+\slash v}{2}\left(-P_5\gamma_5-i\slash P\right)\
,\label{h}\ee and $P_5$, $P^\mu$ are annihilation operators of the
pseudoscalar $P$
 and vector $V$ heavy mesons  normalized as follows:
 \be
 \langle 0|P_5|P\rangle =\sqrt{M_H}~,~~~~~~~~~
 \langle 0|P_\mu|V(\epsilon)\rangle =\epsilon_\mu \sqrt{M_H}
 \ee
The first term in the lagrangian (\ref{wise}) contains the kinetic
term for the heavy mesons giving the $P$ and $P^*$ propagators, $
i/(2v\cdot k)$ and $- i(g^{\mu\nu}-v^\mu v^\nu)/(2v\cdot k)$
respectively. However, since the residual momentum is not small we
will use the complete form of the propagators instead of their
approximate expressions \cite{Isola:2001ar,Isola:2001bn}. The
interactions among heavy and light mesons are obtained by the
other three terms.  In the second term there are interactions
among the heavy mesons and an odd number of pions coming from the
expansion $\AA_\mu$ . Here
$\AA_\mu=\frac{1}{2}\left(\xi^\dagger\partial_\mu \xi - \xi
\partial_\mu \xi^\dagger\right)$. Expanding the field $\xi=\exp(i{
\pi}/f)$ and taking the traces it provides the coupling in the
lower vertices of Fig. \ref{fig:DDKpi}. The last two terms give
the upper  vertices of Fig. \ref{fig:DDKpi}.  The octet of vector
resonances ($\rho$, $K^*$, etc.) is introduced as the gauge
multiplet according to the hidden gauge symmetry approach of Ref.
\cite{Bando:1988br}. We put
\be \rho_\mu=i\frac{g_V}{\sqrt{2}}\hat\rho_\mu \label{h9} \ee
where $\hat\rho$ is the usual hermitian $3\times 3$ matrix of
flavor $SU(3)$ comprising the nonet of light vector mesons and
$g_V$ is determined by vector meson dominance as follows:
$g_V\simeq 5.8$ \cite{Bando:1988br}.

Also the parameter $\beta$ can be fixed by vector meson dominance.
This corresponds to assume that, for the heavy  pseudoscalar
mesons $H$, the coupling of the electromagnetic current to the
light quarks  is dominated by the $\rho,\,\omega,\,\phi$ vector
mesons. This produces the numerical result \be
\beta=\frac{\sqrt{2}m_V}{g_V f_V}\approx 0.9\label{14} \ .\ee

For $g$ we take  the recent experimental result from the CLEO
collaboration, obtained by the full width of $D^{*+}$
\cite{Ahmed:2001ab}; this gives a value: $g=0.59\pm 0.07\pm 0.01$.

It can be noted that while these determinations consider soft
pions we are interested in the coupling of a hard pion to $D$ and
$D^*$. This introduces a correction that can be parametrized by a
form factor, see below for a discussion.

 Let us now consider the parameter $\lambda$ in Eq. (\ref{wise}). In
 \cite{Casalbuoni:1996pg} one can found an estimate
  of this parameter based on the  effective chiral lagrangian for heavy mesons.
  We  present here an update of this analysis and new numerical results.

It is useful to begin by writing down  the phenomenological
heavy-to-heavy  current in the leading $1/m_Q$ approximation:
\be J_{\mu}^{cb} = -\xi (v \cdot v' ) < {\bar H}^{(c)}_a
\gamma_{\mu} (1 - \gamma_5 ) H^{(b)}_a >\ , \label{effcurrf} \ee
which gives rise to \bea <D(v')|{\bar c} \gamma_\mu b |B(v)> & =&
\sqrt{M_B M_D}\, \xi (v \cdot v')\, (v+v')_\mu\ ,
\nn \\
<D^* (v', \epsilon)|{\bar c} \gamma_\mu b |B(v)> & =& \sqrt{M_B
M_D}\,  \xi (v \cdot v')\, \epsilon_{\mu\nu\alpha\beta}
\epsilon^{*\nu} v^{\alpha} v'^\beta\ ,
\nn \\
<D^* (v', \epsilon)|{\bar c} \gamma_\mu \gamma_5 b |B(v)> & =&
\sqrt{M_B M_D}\, \xi (v \cdot v')\, i\, \left[ (1 + v \cdot v')
\epsilon^*_\mu - (\epsilon^* \cdot v) v'_\mu \right] \; ,
\label{wisgur} \eea
where $\xi (v \cdot v' )$ is the Isgur-Wise form factor. Note
that, by this definition, the choice of phases in Eq.
(\ref{wisgur}) agrees with (\ref{v1}) and (\ref{v2}). We also
write the phenomenological heavy-to-light leading current for
coupling to light pseudoscalar mesons of the heavy mesons in the
multiplet $H$
\be
 L^{\mu}_a = \frac{ i {\hat F}}{2} <\gamma^{\mu}
( 1 - \gamma_5 )H_b \xi^{\dagger}_{ba}>~~. \label{c1} \ee
Here $\hat F$ is related to leptonic decay constant of the
pseudoscalar heavy meson $B$ appearing in (\ref{leptonic}) by
\be f_{B} = \frac{{\hat F}}{\sqrt{M_B}}~~. \label{c3} \ee
Several numerical analyses based on QCD sum rules have been
performed, see for a review e.g. \cite{Casalbuoni:1996pg}. The
result we shall use is
\be {\hat F}= 0.30 \pm 0.05\ {\rm GeV}^{3/2} \label{eq:eqf} \ee
which is obtained neglecting radiative corrections since they are
neglected also in the evaluation of the Isgur-Wise form factor the
parameterization we use below is based on.

The effective theory approach gives predictions for the form
factor $V(q^2)$ at high $q^2$ and relates it to the parameter
$\lambda$. We can match this result with the theoretical
calculations coming from Light Cone Sum Rules \cite{Ball:1998kk},
\cite{Colangelo:2000dp} (LCSR) and lattice QCD
\cite{Flynn:1996dc,Burford:1995fc,DelDebbio:1998kr}. To do this we
compute the form factor at $q^2\simeq q^2_{max}$ where it is
dominated by the nearest low-lying vector meson pole. Computing
the form factor at $q^2 \simeq q^2_{max}\equiv (M_B-M_V)^2$ and at
leading order in $1/m_Q$, one gets \cite{Casalbuoni:1992dx}: \be
\,V(q^2_{\rm max})=\frac {g_V} {\sqrt{2}} \lambda {\hat F} \frac
{M_P + M_{V}}{\sqrt{M_P}}\frac{1}{M_V + \Delta} \label{v14} \ee
 where $\Delta$ is the  appropriate mass splitting, i.e., for the
transition $B\to K^\star$, $\Delta=m_{B^*_s}-m_B \simeq 135$ MeV.
From the LCSR and lattice QCD analyses for the transition $B\to
K^*$ we infer the value $V= 1.5$ at $q^2\equiv \bar q^2=17$
GeV$^2$. In order to compare with Eq. (\ref{v14}) we assume that
in the range $q^2\in (\bar q^2,q^2_{max})$ the form factors are
dominated by the $B_s^*$ simple pole. This gives $V(q^2_{max})=
1.8$ and, as a result, \be \lambda=\,+\,0.56~~{\rm GeV}^{-1}~. \ee
We note that to determine $\lambda$ we used a positive sign of
$V(q^2_{max})$. The same sign would be obtained using HQET and by
assuming that the strange mass as large, see  eq. (\ref{wisgur}).
We also note that a different result for the phase and the
magnitude of $\lambda$ was obtained in \cite{Casalbuoni:1996pg}.
The difference in magnitude was due to the different methods used.
Here we use LCSR and lattice QCD, while in
\cite{Casalbuoni:1996pg} an extrapolation is used from the $D\to
K^*$ form factor. As to the phase, it is due to a different choice
we use here for the phase factor of the heavy vector meson, see
eq. (\ref{h}).

To compute the matrix elements we use the following kinematics:
\begin{equation}
p^\mu~=~m_B v^\mu=(m_B,\vec 0 )\ ,\ ~~~~
p^\mu_{D^{(*)}}~=~m_{D}v^{\prime\mu}\ ,\ ~~~~ q~=~p-p_{D^{(*)}}\ .
\end{equation}
where $\dd \omega^* \ =\ \frac{m_B^2+m_D^2-m^2_{D_s}}{2 m_D m_B}$
and the angular integration is over the directions of the vector
$\vec v^{\,\prime}=\vec n \sqrt{\omega^{^* 2}-1}$. Our results are
as follows:
\begin{eqnarray} {\cal M}(B(v)\to D_s(q) D^{*}(\epsilon, v^\prime))&=&
 - K
 \,(m_B+m_D)\,\epsilon^{*}\cdot v\,,\\
{\cal M}(B(v)\to D_s^{*}(\eta ,q) D(v^\prime))&=&  - K\,m_{D_s}\,
\eta^*\cdot(v+v^\prime)\,, \\
 {\cal M}(B(v)\to D_s^{*}( \eta, \, q )D^{*}( \epsilon, v^\prime ))&=&  -\,i  K\,
m_{D_s}\eta^{*\mu}\epsilon^{*\alpha}\left(i\epsilon_{\alpha\lambda\mu\sigma}
v^{\prime\lambda}v^\sigma-g_{\mu\alpha}(1+\omega^*)+v_\alpha
v_{\mu}^\prime \right)\ ,
\end{eqnarray}
where $\dd K\ =\ \frac{G_{F}}{\sqrt 2}V_{cb}^* V_{cs}\ a_2
\sqrt{m_B m_D} f_{D_s}\xi_{IW}(\omega^*)\ .$ On the other hand,
for the $K^*\pi$ final state we have
\begin{eqnarray}
&&{\cal M}(D_s(q)\, D^{*}(\epsilon, v^\prime)\to
K^*(p_K,\hat\epsilon)\pi(p_\pi))= \,  -\, \frac{2 g F^2(|\vec
p_\pi|) }{f_\pi}\,\frac{g_V}{\sqrt 2}\sqrt{
\frac{m_{D^*}}{m_{D_s}}}\,\epsilon_\lambda\hat\epsilon^*_\sigma\cr&&\cr
&&\times\left[\frac{2\beta\,m_D\, q^\sigma p_\pi^\lambda} {
(m_{D}v^\prime-p_\pi)^2-m^2_{D} }\ +\
\frac{4\,\lambda\,m_{D^*}\,G^{\sigma\lambda}(p_\pi,p_K,
v^\prime)}{ (m_{D^*}v^\prime-p_\pi)^2-m^2_{D^*}
}\right]\,,~~~~~~~~~~~~ \label{eq:ab}
\\&&\cr&&\cr &&{\cal M}(D_s^{*}(\eta ,q) D(v^\prime)\to
K^*(p_K,\hat\epsilon)\pi(p_\pi))= \,  -\,\frac{2 g\,m_{D^*}\,
F^2(|\vec p_\pi|) }{f_\pi}\frac{g_V}{\sqrt 2}\sqrt{
\frac{m_{D}}{m_{D_s}}}\,\frac{\eta_\lambda\hat\epsilon^*_\sigma}{
(m_{D^*}v^\prime-p_\pi)^2-m^2_{D^*} }\cr&&\cr &&\times\left[2\beta
q^\sigma\left(+p_\pi^\lambda\, -\,\frac{v^\prime\cdot
p_\pi}{m_{D^*}} \,p_K^\lambda \right)-4\lambda
m_{D_s}H^{\sigma\lambda}(p_\pi,p_K, v^\prime)
\right]\,,\label{eq:c}\end{eqnarray}

\begin{eqnarray}
&&{\cal M}(D_s^{*}(\eta ,q) D^{*}(\epsilon, v^\prime) \to
K^*(p_K,\hat\epsilon)\pi(p_\pi))= \, + \, \frac{2 g\,m_{D^*}\,
F^2(|\vec p_\pi|) }{f_\pi}\frac{g_V}{\sqrt 2}\sqrt{
\frac{m_{D^*}}{m_{D^*_s}}}\,\epsilon^{\alpha\mu\nu\lambda}\,\eta_\tau
\hat\epsilon^*_\sigma\epsilon_\rho \cr &&\cr &&\times\Big[
\frac{4\lambda \,m_D q_\alpha (p_K)_\mu
p_\pi^\rho}{(m_{D}v^\prime-p_\pi)^2-m^2_{D}}
\frac{\delta^\sigma_\nu\delta^\tau_\lambda}{m_{D^*}}\,+\,
\frac{v^\prime_\alpha (p_\pi)_\mu\delta^\rho_\nu
}{(m_{D^*}v^\prime-p_\pi)^2-m^2_{D^*}}\left( 2\beta
q^\sigma\delta^\tau_\lambda\,+\,4\lambda
m_{D^*_s}(p_K^\tau\delta^\sigma_\lambda-(p_K)_\lambda
g^{\sigma\tau})\right) \Big]\, ,\label{eq:de}
\end{eqnarray}
where
\begin{eqnarray} G^{\sigma\lambda}(p_\pi,p_K, v^\prime)= &-& (v^\prime\cdot q)
\biggl(g^{\sigma\lambda}(p_K\cdot p_\pi) -p_\pi^\sigma
p_K^\lambda\biggr)\ -\ (q\cdot
p_\pi)\biggl(v^{\prime\sigma}p_K^\lambda-g^{\sigma\lambda}(v^\prime\cdot
p_K)\biggr)\cr &&\cr &-& q^\lambda\biggl(p_\pi^\sigma (p_K\cdot
v^\prime) - v^{\prime\sigma} (p_K\cdot p_\pi)\biggr)\ ,
\end{eqnarray}
\begin{eqnarray}
H^{\sigma\lambda}(p_\pi,p_K, v^\prime)&= &g^{\sigma\lambda}
\left(p_K\cdot p_\pi\,-\, \frac{v^\prime\cdot
p_\pi}{m_{D^*}}(m_{K^*}^2-p_K\cdot
q)\right)-p_K^\lambda\left(p_\pi^\sigma\,  +\,\frac{v^\prime\cdot
p_\pi}{m_{D^*}} \,q^\sigma\right)\ .
\end{eqnarray}
Eq. (\ref{eq:ab}) corresponds to the sum of diagrams (a) and (b)
of Fig. \ref{fig:DDKpi}; Eq. (\ref{eq:c}) corresponds to diagram
(c) and, finally, Eq. (\ref{eq:de}) corresponds to the sum of
diagrams (d) and (e). Similar results hold for the
$K(\rho,\omega,\phi)$ final states and we do not reproduce them
here for brevity sake. $F(|\vec p_\pi|)$ is a form factor taking
into account that in the vertex  $DD^* \pi$ the pion is not soft
and therefore the coupling constant should be corrected. Its
determination by a quark potential model is discussed in
\cite{Isola:2001ar} using a quark model. The central value we use
is $F(|\vec p_\pi|)\, = \,0.065$. In absence of more detailed
information, the same form factor is adopted for the upper
vertices in Fig. \ref{fig:DDKpi} as well.

\section{Numerical results\label{sec.4}}
Let us first consider the short distance contribution. Using, for
the CKM matrix elements, $\rho = 0.2296$, $\eta = 0.3249$, $A =
0.819$, and for the involved form factors the value collected in
Table IV of Ref. \cite{Ali:1998nh}, one  gets the results reported
in Tab. \ref{tab:2}. These values correspond to the following
value of the angle $\gamma$ of the unitarity triangle: 
\be \gamma= arctan\left( \frac{\eta}{\rho}\right) \simeq
54.8^\circ\,. \label{ga} \ee 
\begin{table}[ht!]
\caption{\label{tab:2} {\small  Theoretical values for ${\cal
M_{SD}}$, ${\cal M_{LD}}$. Units are GeV.\Black  }}
\begin{center}
\begin{tabular}{|c|c|c|}
\hline {\rm Process}  & ${\cal M_{SD}} \times 10^8$ &  ${\cal
                  {\rm Im}\,M_{LD}}\times 10^8$\\
\hline
$  B^{+}\, \to  K^{\ast 0} \pi^+$ & $~ +1.45~$ & $ -2.14 ~ $\\
\hline
$  B^{+}\, \to K^{\ast +} \pi^0$ & $~ +1.02 - 0.79 \,i  ~$  & $-1.52  ~ $ \\
\hline
$ B^0\, \to K^{\ast 0} \pi^0$ & $~-0.60 - 0.08\,i ~$ & $ +1.52 ~$\\
\hline
$ B^0 \, \to  K^{\ast +} \pi^-$& $~+0.85 - 1.01\,i ~$ & $ -2.14~$ \\
\hline
$  B^{+}\, \to K^{0} \rho^ +$  & $+0.17~$ & $-2.74 $\\
\hline
$B^{+}\, \to K^{+}\rho^0$& $~+0.39 - 0.64\,i  ~$  & $ -1.94$ \\
\hline
$B^0\, \to K^{0} \rho^0$& $~+0.48- 0.11\,i  ~$  & $+1.94~$\\
\hline
$ B^0\, \to K^{+} \rho^-$& $~-0.28 - 0.74\,i  ~$  & $-2.74  $ \\
\hline
$ B^{+}\, \to K^{+}\omega $ & $~-0.27 - 0.63\,i $  & $ -1.94  $ \\
\hline
$ B^{0}\, \to K^{0} \omega$& $~+0.03 - 0.10\,i $  & $ -1.94 $\\
\hline
$ B^{+}\, \to K^{+}\phi $ & $~1.30 $  & $-2.58$ \\
\hline
$ B^{0}\, \to K^{0}\phi $ & $~1.30 $  & $-2.58$ \\
\hline
\end{tabular}
\end{center}
\end{table}
Next we consider the long distance absorptive part. For its
parameters  we assume the  values of previous Section. For the
$D_s$ decay constant we use $f_{D_s}= {\hat F}/\sqrt{m_{D_s}}$,
with $\hat F$ given in (\ref{eq:eqf}); for the Isgur-Wise form
factor we use the parameterization
\be \xi(v\cdot v')=\left(\frac{2}{1+v\cdot v'}\right)^{\hat
\rho^2} \ee
with $\hat \rho^2\simeq 1$. This parametrization agrees with the
results for the $b\to c$ exclusive decays obtained by the CLEO and
Belle Collaboration. It is a fit to the results obtained by the
QCD sum rules method, see \cite{Casalbuoni:1996pg}.

The numerical results for the imaginary part of ${\cal M_{LD}}$
are given in Tab.
   \ref{tab:2}. Typical sizes
  of the different contributions to
  $  {B}^{-}\, \to \overline{ K}^{*0} \pi^-$ are as follows.
   From the two terms in Eq.(\ref{eq:ab}):
    ${\rm Im }{\cal M}^{a,b}_{\cal LD}= -1.32 \times 10^{-8}$; from
the  terms  in Eq.(\ref{eq:c}):   ${\rm Im }\,{\cal M}^{c}_{\cal
LD}=-0.16\times 10^{-8}$;  from the  terms in Eq.(\ref{eq:de}):
${\rm Im }\,{\cal M}^{d,e}_{\cal LD}= -0.66\times 10^{-8}$. For
the
 $  {B}^{-}\, \to { K^{0}} \rho^-$
   channel we find: ${\rm Im }\,{\cal M}^{a,b}_{\cal LD}= -1.52
    \times 10^{-8}\,$;   ${\rm Im }{\cal M}^{c}_{\cal
LD}=-0.70\times 10^{-8}\,$; ${\rm Im }{\cal M}^{d,e}_{\cal LD}=
-0.52\times10^{-8}\,$. It can be noted that the phase of
 ${\cal M_{SD}}$ is purely weak while the phase in ${\cal M_{LD}}$
 is only due to strong interactions.

\subsection{Branching ratios}
From the results of Tab. \ref{tab:2} we can compute the Branching
Ratios (${\cal B}$) and the CP asymmetries. As explained in the
Introduction, we have not presented an evaluation of the real part
of ${\cal A_{LD}} $. Attempts for the $K\pi$ channel can be found
in \cite{Isola:2001ar,Isola:2001bn}. Typical results are that the
real part is of the same order of the imaginary part, but the
uncertainties are large. To estimate the CP averaged Branching
Ratios we add therefore to the imaginary part a real part as
follows: \be {\cal M_{LD}}= {\cal R_{LD}}\,+\,i\,{\cal I_{LD}}
\label{mld} \ee and consider the values ${\cal R_{LD}}=0,\,\pm
\,{\cal I_{LD}}$. The results are reported in Tab. \ref{tab:2.5}
for the three values of ${\cal R_{LD}}$, together with the results
obtained considering only the short distance amplitude. They show
the relevance of the long-distance contribution.
\begin{table}[ht!]
\caption{{\small \label{tab:2.5} Branching Ratios ${\cal B}$
(units $10^{-6}$). In the second and  third columns, theoretical
values computed using only the short distance amplitude and  the
full amplitude (i.e. short distance and long distance); in the
latter case the three values  correspond respectively to ${\cal
R_{LD}}=(+1,\,0,\, -1)\,\times\, {\cal I_{LD}}$, see text.}}
\begin{center}
\begin{tabular}{|c|c|c|}
\hline
{\rm Process}   &${\cal B}$ (${\cal SD}$~{\rm only})  & ${\cal B}$ (${\cal SD}$+${\cal LD}$)    \\
\hline
$  B^{+}\, \to  K^{\ast 0} \pi^+$ & $~1.96~$ & $(4.71, 6.22, 16.3)$\\
\hline
$  B^{+}\, \to K^{\ast +} \pi^0$ & $~1.56~$  & $(5.20, 5.95, 11.0)$ \\
\hline
$ B^0\, \to K^{\ast 0} \pi^0$ & $~0.32~$ & $(2.51,2.10,5.66)$\\
\hline
$ B^0 \, \to  K^{\ast +} \pi^-$& $~1.50~$ & $(9.94,9.13,16.2)$ \\
\hline
$  B^{+}\, \to K^{0} \rho^ +$  & $0.03~$ & $(13.1,6.99,14.8) $\\
\hline
$B^{+}\, \to K^{+}\rho^0$& $~0.52 ~$  & $(8.42,6.32,11.2)$ \\
\hline
$B^0\, \to K^{0} \rho^0$& $~0.21~$  & $(7.88, 3.07, 4.71)$\\
\hline
$ B^0\, \to K^{+} \rho^-$& $~0.54~$  & $(18.1,10.4,15.6)$ \\
\hline
$ B^{+}\, \to K^{+}\omega $ & $~0.44$  & $(10.7,6.21,8.73)$ \\
\hline
$ B^0\, \to K^{0} \omega$& $~0.01$  & $(6.70,3.58,6.91)$\\
\hline
$ B^{+}\, \to K^{+}\phi$ & $~1.55$  & $(7.58,7.63,19.8)$ \\
\hline
$ B^{0}\, \to K^{0}\phi$ & $~1.42$  & $(6.98,7.02,18.3)$ \\
\hline
\end{tabular}
\end{center}
\end{table}
The ''best value" of ${\cal R_{LD}}$ might be obtained by a fit to
the available experimental data; these data are reported in Tab.
\ref{tab:3} and a comparison with the results of the previous
table shows that the preferred values, except for the channels
with $\phi$ in the final state, satisfy  ${\cal R_{LD}}\,\approx
\,-\,{\cal I_{LD}}$. Instead, the decay in $K\,\phi$ prefer ${\cal
R_{LD}}\,\approx \,+\,{\cal I_{LD}}$. To find a "best value" of
${\cal R_{LD}}$ we defined a $\chi^2$ as:
\be \chi^2\equiv \sum_i \left( \frac{ Br^{th}_i-Br_i^{exp} }{
\sigma(Br_i) } \right)^2 \ee
where the $Br_i^{exp}$ are reported in Tab. \ref{tab:3} and the
$\sigma(Br_i)$ are obtained summing quadratically the statistic
and systematic errors in the same table. When the experiment
provides an asymmetric error $^{\sigma_1}_{\sigma_2}$ a
conservative error was assumed:  $\sigma =
Max(\sigma_1,\sigma_2)$. For the decay $ \bar B^0\to \bar K^{\ast
0}\pi^0$ we put $Br^{exp}=0$ and the corresponding error was fixed
to be equal to the experimental upper limit $\sigma = 3.6 \times
10^{-6}$. We have first attempted a fit with only one real
parameter $r = {\cal R_{LD}}/{\cal I_{LD}}$, which corresponds to
the use of the SU(3) nonet symmetry for the light vector mesons.
The minimum value of $\chi^2 (=12.1)$ is obtained for $r=1.207$.
Since however we do not expect the validity of the nonet
hypothesis we have also tried a two parameter fit, corresponding
to the two multiplets (octet+singlet) of the light vector mesons.
This gives as a result ${\cal R_{LD}}=\,-0.954\, {\cal I_{LD}}$
for the ($\rho$, $K^*$, $\omega$) set of particles and two
solutions, ${\cal R_{LD}}=\,-0.201\, {\cal I_{LD}}$ and ${\cal
R_{LD}}=\,+1.21\, {\cal I_{LD}}$, for the $\phi$ with the same
$\chi^2 = 7.8$. The former solution would be preferred because it
represents a less important deviation of the SU(3) nonet symmetry.
Even if the two solutions produce the same values of the Branching
Ratios they would produce different CP asymmetries. The results we
find are in Tab. \ref{tab:3}.
\begin{table}[ht!]
\caption{{\small \label{tab:3} CP averaged Branching Ratios ${\cal
B}$ (units $10^{-6}$). In the second column theoretical values
computed using the present model. In third column and fourth
column theoretical computations based on Ref.
\cite{Aleksan:2003qi}: Scenario (Sc.) 1 refers to  QCD with
factorization and free $\gamma$; Scenario  2 refers to
QCD+Charming penguins with constrained $\gamma$ (see text).
Experimental data are from CLEO
 \cite{Jessop:2000bv,Eckhart:2002mb}, BaBar
 \cite{Aubert:2001ap,Aubert:2001zf}, and
 Belle \cite{Lu:2002qp} or averages from these data.
  }}
  \begin{center}
  \begin{tabular}{|c|c|c|c|c|}
    \hline
{\rm Process}    &{\rm ${\cal B}$ (this paper)} &{\rm ${\cal B}$
(\cite{Aleksan:2003qi}, Sc. 1})&{\rm ${\cal B}$
(\cite{Aleksan:2003qi}, Sc. 2})&  {\rm ${\cal B}$(Exp.)}   \\
\hline
$ {B}^{+}\, \to \,K^{*0}\pi^{+}$& $15.6$ & $ 7.889$&$ 11.080$&$12.1\pm 3.1$~(av.)\\
  \hline
$  {B}^{-}\, \to \,K^{*-}\pi^{0}$ & $8.44 $ &$7.303
 $&$8.292
 $&$7.1^{+11.4}_{-7.1}\pm 1.0\,(<\,31)$~(CLEO)\\
  \hline
$  \bar {B}^0\, \to \,\bar K^{*0}\pi^{0}$& $5.61$ ~& $ $&~$ $&~
$ <\,3.6$~(CLEO)\\
  \hline
$  {B}^0\, \to \, K^{*+}\pi^{-}$& $12.0$ ~& $9.760 $&$10.787
$&$19.3\pm 5.2$~(av.)\\
 \hline $  {B}^{+}\, \to
\,K^{0}\rho^{+}$ & $14.1$ & $7.140 $& $14.006$&\\
\hline $  {B}^{+}\, \to \,K^{+}\rho^{0}$ & $8.56 $ ~&~$1.882
 $&$5.665
 $&$8.9\pm 3.6 $~(av.)\\
  \hline
$  {B}\, \to \,K^{0}\rho^{0}$ & $4.86$ ~&~$5.865 $&$8.893
$&\\
  \hline
$  {B^0}\, \to \, K^{+}\rho^{-}$& $11.6$ ~&~$6.531
 $&$14.304
 $&$
15.9\pm 4.7$~(av.)
\\
\hline $  {B}^+\, \to \,K^{+}\omega$& $6.19$ &$2.398 $ &$6.320 $
&$~9.2^{+2.6}_{-2.3}\pm 1.0$~ (CLEO);$~<4 $ (BaBar);~
$ <7.9$ (Belle)~\\
\hline $  {B}^0\, \to \,K^{0}\omega$& $6.27$ & $2.318 $ &$5.606 $
&$ 6.3\pm 1.8$
~(av.)\\
\hline
$  {B}^+\, \to \,K^{+}\phi$& $9.11$ & $8.941 $ &$9.479 $ &$~8.9\pm 1.2$~(av.)\\
\hline $  {B}^0\, \to \,K^{0}\phi$& $8.39$ & $ 8.360$ &$8.898 $
&$~ 8.7\pm 1.4$
~(av.)\\
  \hline
\end{tabular}
\end{center}
\end{table}
Besides our data, we also present two model calculations presented
in \cite{Aleksan:2003qi}, where an analysis of all the $PV$ (not
only strange) final states is performed. The first model (called
in that paper Scenario 1) contains short distance terms (QCD
factorization) and unconstrained $\gamma$ angle. Other theoretical
determinations based on QCD factorization are in
Ref.~\cite{Du:2002cf}, e.g. ${\cal B}(B^0\to K^+\rho^-)= 12.1 $.
The authors of \cite{Aleksan:2003qi} do not agree with  the
conclusions of Du et al., Ref.~\cite{Du:2002cf}, because,
differently from them, they include the $K^*\pi$ channels. The
second model (Scenario 2) uses QCD factorization and charming
penguins as in the present paper, but differently from this paper,
where only two real free parameters are used, they employ two
universal complex amplitudes multiplied by a computed
Clebsh-Gordan coefficient. It is worthwhile to note that a
significant agreement with the data is obtained with our simple
hypothesis. We also note that the value $\gamma$ of Eq. (\ref{ga})
obtained by a global fit to the CKM matrix \cite{Ciuchini:2000de}
is compatible with the $B\to K^*\pi$ ,$K(\rho,\omega,\phi)$
Branching Ratios only if the charming penguin diagrams are
included.

\subsection{CP asymmetries}
From previous results we can also compute the CP asymmetries  for
the various channels. The Belle Collaboration \cite{Lu:2002qp}
reported the result
\be {\cal A}_{CP}^{ K^-\omega}= \frac{{\cal B}(B^-\to
K^-\omega)-{\cal B}(B^+ \to K^+\omega )} {{\cal B}(B^-\to
K^-\omega )+{\cal B}(B^+\to  K^+\omega  )}\ =-0.21 \pm 0.28\pm
0.03~. \ee
The result we find agrees with the data:
\be {\cal A}_{CP}^{ K^-\omega}=-0.37~~~~~~~~({\rm theory})\ . \ee
We also get
\be {\cal A}_{CP}^{\omega \bar K^0}= \frac{{\cal B}(\bar B^0\to
\omega \bar K^0)-{\cal B}(B^0\to
 \omega K^0 )}{{\cal B}(\bar B^0\to \omega \bar K^0)+{\cal B}(B^0\to
 \omega K^0 )}\ =-0.05\, .
\ee
We can similarly compute the asymmetries for the $K^*\pi$, $K\rho$
channels, defined by 
\begin{eqnarray}
{\cal A}_\pi^{+0}&=&\frac{{\cal B}(B^+\to K^{*+}\pi^0)-{\cal
B}(B^-\to K^{*-}\pi^0)}{{\cal B}(B^+\to K^{*+}\pi^0)+{\cal
B}(B^-\to K^{*-}\pi^0)}\ , \cr && \cr
{\cal A}_\pi^{0+}&=& \frac{{\cal B}(B^+\to K^{*0}\pi^+)-{\cal
B}(B^-\to \overline{K}^{*0}\pi^-)}{{\cal B}(B^+\to
K^{*0}\pi^+)+{\cal B}(B^-\to \overline{K}^{*0}\pi^-)}\ , \cr &&
\cr
{\cal A}^{+-}_\pi&=& \frac{{\cal B}(B^0\to K^{*+}\pi^-)-{\cal
B}(\overline{B}^0\to K^{*-}\pi^+)}{{\cal B}(B^0\to
K^{*+}\pi^-)+{\cal B}(\overline{B}^0\to K^{*-}\pi^+)}\ , \cr &&
\cr
{\cal A}^{00}_\pi&=& \frac{{\cal B}(B^0\to K^{*0}\pi^0)-{\cal
B}(\overline{B}^0\to \overline{K}^{*0}\pi^0)}{{\cal B}(B^0\to
K^{*0}\pi^0)+{\cal B}(\overline{B}^0\to \overline{K}^{*0}\pi^0)}\
, \label{asymm}
\end{eqnarray}
with analogous definitions for the $K\rho$ channel, with  ${\cal
A}_\pi\to{\cal A}_\rho$ and the changes $K^*\to K$ and
$\pi\to\rho$.

\begin{table}[ht!]
\caption{\label{tab:5}{\small  Theoretical values for the
asymmetries; for the definitions see (\ref{asymm}). \Black }}
 \begin{center}
  \begin{tabular}{|c|c|c|c|c|c|c|c|c|}
    \hline {\rm Asymmetry} & ${\cal A}^{+0}_\pi$ & ${\cal
    A}^{0+}_\pi$
    & $ {\cal A}^{+-}_\pi$ &$ {\cal A}^{00}_\pi$ &$  {\cal
    A}^{+0}_\rho$ & ${\cal A}^{0+}_\rho$ & ${\cal A}^{+-}_\rho$ & ${\cal A}^{00}_\rho$ \\
 \hline $A_{CP}$
 & $0.27$ & $0$ & $0.31$  & $-0.04$ & $0.27$ & $0$ &$0.30$ & $-0.08$ \\
\hline
\end{tabular}
\end{center}
\end{table}
The results are presented in Table \ref{tab:5}. They have a
peculiar pattern and will therefore  provide a crucial test of the
present model when future experimental data for these asymmetries
are available.

\vspace{0.8truecm} \textbf{Acknowledgements} We thank T.N.Pham for
precious collaboration at an early stage of this work. C.I. is
partially supported by MIUR under COFIN PRIN-2001.

\bibliographystyle{apsrev}
\bibliography{charmingPV}
\end{document}